\documentclass[onecolumn, aps, prd, groupadress, nofootinbib, superscriptaddress, showkeys, showpacs]{revtex4-1}

\usepackage{amsmath, amsfonts, amssymb, mathrsfs}
\usepackage{amsthm}
\usepackage{tensor}
\usepackage{multirow}
\usepackage{bbm} 
\usepackage[geometry]{ifsym}
\usepackage{placeins}

\usepackage{scalerel,graphicx}
\usepackage[font=small,labelfont=bf]{caption}
\usepackage[font=small,labelfont=bf,textfont=normal,justification=justified,singlelinecheck=false]{subcaption}

\usepackage{cmbright}

\usepackage{graphicx}
\usepackage{float}
\usepackage[font=small]{caption}

\usepackage[hyperindex]{hyperref}
\hypersetup{breaklinks=true, hyperfootnotes=true, pagecolor=white, colorlinks=true}
\usepackage[all]{hypcap}

\makeatletter
\DeclareRobustCommand*{\bfseries}{%
  \not@math@alphabet\bfseries\mathbf
  \fontseries\bfdefault\selectfont
  \boldmath
}
\makeatother

\renewcommand{\nabla}{\!\mathrel{\raisebox{.15em}{%
           \reflectbox{\rotatebox[origin=c]{180}{$\triangle$}}}}\!\!} 

\newcommand{\Dslash}{{D\kern-0.63em{/}}}

\begin{document}

\title{Is the electromagnetic field responsible for the cosmic acceleration in late times?}

\

\author {M. Novello and A. E. S. Hartmann}
 \affiliation{
Centro de Estudos Avan\c{c}ados de Cosmologia/CBPF \\
Rua Dr. Xavier Sigaud 150, Urca 22290-180 Rio de Janeiro, RJ-Brazil}
\date{\today}

\vspace{2 cm}

\date{ \today}

\begin{abstract}
We show that in the realm of general relativity a non minimal coupling between the electromagnetic and the gravitational fields may produce an era of accelerated expansion.
\end{abstract}

\maketitle

\section{Introduction}

It is a common knowledge, described in almost all text book of cosmology that the evolution of the universe may be separated in different eras, according to the dependance of the energy distribution on the expansion factor $ a(t)$ for the geometry of the homogeneous and isotropic metric of the universe. In the present paper we consider that the 3-d space is the Euclidean section that allows to write the metric under the simple form

\begin{equation}
ds^{2} = dt^{2} - a^{2}(t) \, (dx^{2} + dy^{2} + dz^{2})
\label{1}
\end{equation}

In the conventional approach the first era is controlled by the electromagnetic radiation that is followed by a matter dominant era. This description is related to the conservation of the different perfect fluids that may control the expansion of the universe. Thus, the density of radiation goes like $ a^{- 4} $ and the energy density of the pressureless mater depends  on $ a^{- 3}.$ It is then understood that there is no more important influence of the electromagnetic field in the expansion of the universe beyond such radiation era. We will argue that it may not be the case and the importance of the electromagnetic field can go far beyond such prescription.

The recent proposals of describing matter in terms of an unknown dark energy provides a challenge to this simple schema, once it was suggested that new forms of strange matter may be the responsible for a new era that instead of decelerating the universe at its expands, what should be interpreted as the most natural circumstance, provokes an unexpected acceleration. Many proposals to solve this situation have been investigated \cite{de}.

In this paper we follow another path and ask if it should be possible that instead of requiring new forms of matter it should not be the case of conventional matter to do the job. Indeed, we shall prove that it is possible that a combination of the electromagnetic and the gravitational field in a non minimal interaction could be responsible for a late acceleration. In the realm of General Relativity such interaction can be interpreted in terms of an extra energy momentum tensor. Besides, we shall see that in an expanding universe with Euclidean section this extra term is nothing but a perfect fluid. Then we show that the combination of the energy-momentum tensor of the minimal coupling term added to the energy-momentum tensor of the non minimal part can generate  an acceleration of the scale factor in a later epoch of the expanding universe. On the other hand for primordial times the behaviour of the electromagnetic fluid is the standard one described in the minimally coupling case. In other words, the important modification occurs for late times.

\section{The field equations}

We start by considering a combination of Einstein \rq s general relativity with Maxwell electrodynamics plus an extra term of non minimal coupling \cite{nb} through the action principle

\begin{equation}
\delta\int \, \sqrt{-g} \,  \left(\frac{1}{2 \,k} \, R - \frac{c^{2}}{4} \, F + \frac{\xi}{2} \, R \, F \right)= 0
\label{14jan1}
\end{equation}
where $ F = F_{\mu\nu} \, F^{\mu\nu}.$

 The equation of the metric is given by

\begin{equation}
  R_{\mu\nu} - \frac{1}{2} \, R \, g_{\mu\nu} = - \, T_{\mu\nu}^{\gamma} - \,\xi \,  Z_{\mu\nu} = - T_{\mu\nu}
  \label{gr1}
  \end{equation}
The Maxwell energy-momentum tensor is the standard form
$$ T_{\mu\nu}^{\gamma} = \Phi_{\mu\nu} + \frac{1}{4} \, F \, g_{\mu\nu}$$
where $ \Phi_{\mu\nu} = F_{\mu\alpha} \, F^{\alpha}{}_{\nu}.$ The extra non minimal term $Z_{\mu\nu}$ is

\begin{equation}
Z_{\mu\nu} = F \, ( R_{\mu\nu} - \frac{1}{2} \,R \,  g_{\mu\nu}) - 2 \, R \, \Phi_{\mu\nu} - \Box F \, g_{\mu\nu} + F_{;\mu ;\nu}
\end{equation}
 We set from now on $ k =1 = c.$

The equation for the electromagnetic field is

$$ F^{\mu\nu}{}_{; \nu} - 2\, \xi \, ( R \, F^{\mu\nu} )_{; \nu} = 0.$$

\section{the average procedure: $ Z_{\mu\nu}$ interpreted in terms of a fluid}

In the case of non minimal coupling with gravity the extra term for the energy-momentum tensor $ Z_{\mu\nu}$ is rather involved once it contains terms depending on the curvature. Let us analyze the case in which the metric takes the form (\ref{1}).

In this case
$$ R_{00} = \dot{\theta} + \frac{1}{3} \, \theta^{2}, $$
$$ R_{ij} = \frac{1}{3} \,( \dot{\theta} +  \, \theta^{2}) \, g_{ij} $$
$$ R = 2 \, \dot{\theta} +  \frac{4}{3}\, \, \theta^{2}$$ where we have defined the expansion factor $ \theta = 3 \, \dot{a}/a.$

We are interested in analyze the effects of the non minimal coupling in the standard spatially isotropic and homogeneous metric. To be consistent with the symmetries of this choice of the metric,
an averaging procedure must be performed if electromagnetic fields
are to be taken as a source for the gravitational field, according to the standard procedure \cite{tolmanbook}. As a consequence, the components of the electric
$E_{i}$ and magnetic $B_{i}$ fields must satisfy the following
relations:
\begin{eqnarray}
<{E}_i > = 0,\qquad
%
<{B}_i> &=& 0,\qquad
%
<{E}_i\, {B}_j> = 0,
\label{meanEH}\\[1ex]
<{E}_i\,{E}_j> &=& -\, \frac{1}{3} {E}^2
\,g_{ij},
\label{meanE2}\\[1ex]
<{B}_i\, {B}_j> &=&  -\, \frac{1}{3} {B}^2
\,g_{ij}.
\label{meanH2}
\end{eqnarray}
where $E$ and $B$ depends only on time.

Using the above average values it follows that the Maxwell energy-momentum tensor $T_{\mu\nu}^{\gamma}$
reduces to a perfect fluid configuration with energy density
$\rho_\gamma$ and pressure $p_\gamma$ given by
\begin{equation}
<T_{\mu\nu}^{\gamma} > = (\rho_\gamma + p_\gamma)\,
V_{\mu}\, V_{\nu} - p_\gamma\, g_{\mu\nu}, \label{Pfluid}
\end{equation}
where
\begin{equation}
\label{RhoMaxwell} \rho_\gamma = 3p_\gamma = \frac{1}{2}\,(E^2 +
{B}^2) = \frac{(\sigma^{2} + 1)}{2} \, X
\end{equation}
once
$$ < \Phi^{\mu\nu}> = \frac{2}{3} \, (\sigma^{2} + 1) \, X \, V_{\mu} \, V_{\nu} + \frac{1}{3} \, (\sigma^{2} - 2) \, X \, g_{\mu\nu} $$
where we have set
$ E^{2} = \sigma^{2} \, B^{2}$ and $ V_{\mu} = \delta_{\mu}^{0}.$    For simplicity, from now on we will write $ B^{2} = X.$ Let us decompose the extra term $  < Z_{\mu\nu} > $ as a fluid. The average tensor $ Z_{\mu\nu} $ is given by

\begin{eqnarray}
< Z_{\mu\nu} > &=& F \, ( R_{\mu\nu} - \frac{1}{2} \, R \, g_{\mu\nu} ) - \frac{4}{3} ( \sigma^{2} + 1) \, X \, R \, V_{\mu} \, V_{\nu} + \nonumber \\
&-& \frac{2}{3} \,(\sigma^{2} - 2) \,  X \, R \, g_{\mu\nu} -  \Box F \, g_{\mu\nu} + F_{; \mu ; \nu}.
\end{eqnarray}

where $ F = - \, 2 \, (\sigma^{2} - 1 ) \, X.$

\subsection{The perfect fluid}

From what we have shown we can analyze the effects of the non minimal coupling of the electromagnetic field with gravity in a spatially homogeneous and isotropic geometry in terms of the standard equation of general relativity and a mixed fluid as in equation (\ref{gr1}). In order to compatibilize the theory with a spatially homogeneous and isotropic metric one must analyze its corresponding heat flux $ q_{\mu}$ and the anisotropic pressure $ \pi_{\mu\nu}$ which are defined for an arbitrary energy-momentum tensor $T_{\mu\nu}$ as

  $$ q_{\lambda} = T_{\alpha\beta} \, V^{\beta} \, h^{\alpha}{}_{\lambda} $$

  $$ \Pi_{\mu\nu} = T^{\alpha\beta} \, h_{\alpha\mu} \, h_{\beta\nu} + p \, h_{\mu\nu}$$
  where
  $$ h_{\mu\nu} = g_{\mu\nu} - V_{\mu} \, V_{\nu} $$ and the pressure
  $$ p = - \, \frac{1}{3} \, T_{\mu\nu} \, h^{\mu\nu}.$$

In the case of tensor $ Z_{\mu\nu}$ a direct calculation shows that both  $ q_{\mu}$ and $ \pi_{\mu\nu}$ vanish. This yields the remarkable consequence that the non minimal coupling part of the energy-momentum tensor can be interpreted in terms of a perfect fluid too, like in the minimal coupling case. That is we can write the coupled energy-momentum tensor $ Z_{\mu\nu}$ in the form of a perfect fluid
$$ Z_{\mu\nu} = (\rho_{Z} + p_{Z} ) \, V_{\mu} \, V_{\nu} - p_{Z} \, g_{\mu\nu} $$
where
\begin{equation}
\rho_{Z}= 2 \, (\sigma^{2} - 1 ) \, \theta \, \dot{X} - 4 \, \sigma^{2} \, \dot{\theta} \, X - 2 \, ( \sigma^{2} + \frac{1}{3} ) \, \theta^{2} \, X.
\label{25121}
\end{equation}
and for the pressure we find

\begin{eqnarray}
p_{Z}&=& - \, 2 \, (\sigma^{2} - 1 ) \, \ddot{X} - \frac{4}{3} \, ( \sigma^{2} - 1) \, \theta \, \dot{X} \nonumber \\
&-& \frac{4}{3} \, \dot{\theta} \, X + \frac{2}{9} \, ( \sigma^{2} - 5 ) \, \theta^{2} \, X.
\label{25122}
\end{eqnarray}

\section{The case $ \sigma^{2} = 0:$ the magnetic universe}

In order to simplify our exposition let us consider the particular case of a magnetic universe by setting $ \sigma^{2} = 0.$

Thus the total energy and the total pressure to be used as source of gravity in equation (\ref{gr1})
$$ T_{\mu\nu} = (\rho + p) \, V_{\mu}\, V_{\nu} - p \, g_{\mu\nu} $$
are provided by the quantities

$$ \rho = \rho_{\gamma} + \xi \, \rho_{Z} $$
$$ p = p_{\gamma} + \xi \, p_{Z}$$

that is
\begin{equation}
\rho = \frac{1}{2} \, X - 2 \, \xi \, \theta \, \dot{X} - \frac{2}{3} \xi \,  \theta^{2} \, X .
\label{15021}
\end{equation}

\begin{equation}
p = \frac{1}{6} \, X + 2 \, \xi \, \ddot{X} + \frac{4}{3}  \, \xi \, \theta \, \dot{X} -  \frac{4}{3}  \, \xi \, \dot{\theta} \,X  - \frac{10}{9} \, \xi \, \theta^{2} \, X.
\label{15022}
\end{equation}

The unique equations that are not trivially satisfied are the $ 0-0$ components and the conservation of the total energy, that is

\begin{equation}
 \rho = \frac{1}{3} \, \theta^{2}
 \label{221}
 \end{equation}

 \begin{equation}
\dot{\rho} + (\rho + p) \, \theta = 0
\label{223}
\end{equation}

Using the above expressions for $ \rho$ and $ p$ the equation of conservation reduces to
\begin{equation}
\left( \frac{1}{2} \, \dot{X} + \frac{2}{3} \, \theta \, X \right) \, \left(1 - 4 \,\xi \, \dot{\theta}  - \frac{8}{3} \, \xi \theta^{2}\right) = 0
\label{212}
\end{equation}
This is certainly an unexpected and  remarkable result once it implies that the equation of the conservation law can be factored into two independent ones and must be examined separately . We concentrate our attention to the equation

\begin{equation}
 1 - 4 \,\xi \, \dot{\theta}  - \frac{8}{3} \, \xi \theta^{2} = 0
 \label{231}
 \end{equation}
once the other path is uninteresting for our discussion.
 This case is the most relevant for our interest once it provides the evolution of the expansion factor $ \theta$ that can be obtained immediately by a direct integration. We obtain

\begin{equation}
\theta = \frac{3}{4} \, m \, \frac{(e^{mt} + 1)}{(e^{mt} - 1)}
\label{213}
\end{equation}

The equation for $X$ is provided by $ \rho = \theta^{2}/3$  that is

$$ \dot{X} + f(t) \, X = g(t).$$
where
$$ f(t) = \frac{1}{3} \,\theta - \frac{1}{4 \, \xi \, \theta} $$

$$ g(t) = - \, \frac{1}{6 \, \xi} \, \theta.$$

A direct integration yields

\begin{equation}
X(t) = \frac{3}{4} m^{2} \, \frac{(1 + e^{mt})}{(1 -  e^{mt})} \, F(1/4, 1/2, 5/4; e^{mt})
\end{equation}

where $ F(1/4, 1/2, 5/4; e^{mt}) $ is the hypergeometric function and the constant $ m$ is related to the constant of interaction $ \xi $ by the relation

$$ m^{2} = \frac{2}{3\,\xi}. $$

\section{Late times acceleration}

Let us consider the case in which the constant  $m $ is negative. Set $ m = - \, \beta^{2}.$ Thus we have for the scale factor $ a(t) $ and the expansion $ \theta$ the values \cite{anysigma}

$$ a(t) =e^{-\beta^{2}\, t/4} \, \sqrt{e^{\beta^{2}\, t} - 1} $$

$$\theta = \frac{3 \, \beta^{2}}{4} \, \frac{(e^{\beta^{2}\, t} + 1)}{(e^{\beta^{2}\, t} - 1)}.$$

In the limit of big values of t the acceleration becomes positive

$$ \lim_{t \longrightarrow \infty}\, \frac{\ddot{a}}{a} = \frac{1}{16} \, \beta^{4} = \frac{\theta_{\infty}^{2}}{9}.$$

On the other hand, for primordial times, when $ \beta^{2} \, t << 1$ the system behaves as the electromagnetic radiation in the case of minimally coupling to gravity, that is

$$ a(t) \approx \sqrt{t} $$

$$ \theta \approx  \frac{3}{2 \, t}.$$

\begin{figure*}[!htb]
\centering
\caption{Magnetic case: evolution of $a(t)$ for $m<0$.}
\includegraphics[width=.45\textwidth]{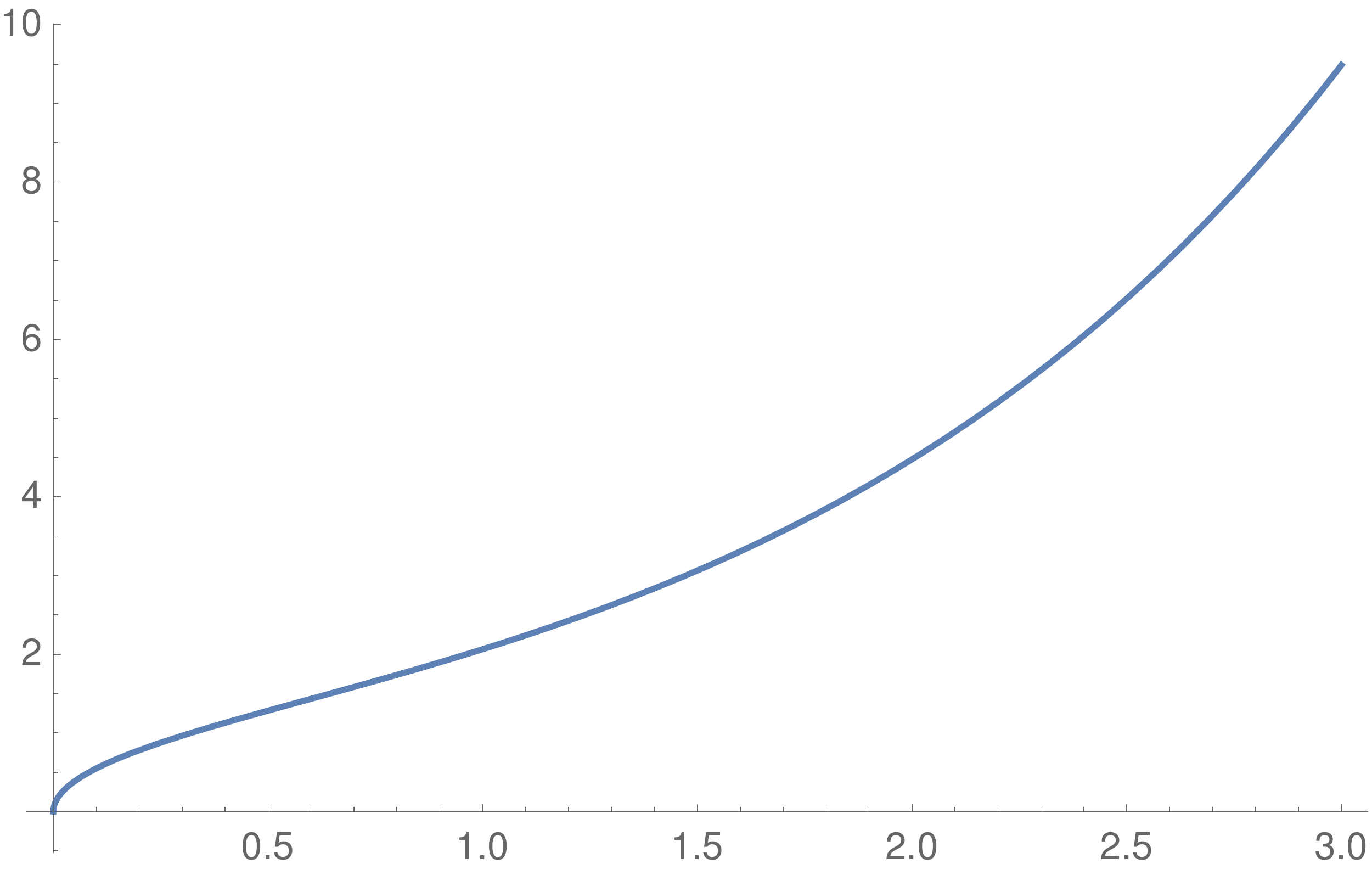}
\end{figure*}

\begin{figure*}[!htb]
\centering
\caption{Magnetic case: solution for $X(t)$.}
\includegraphics[width=.45\textwidth]{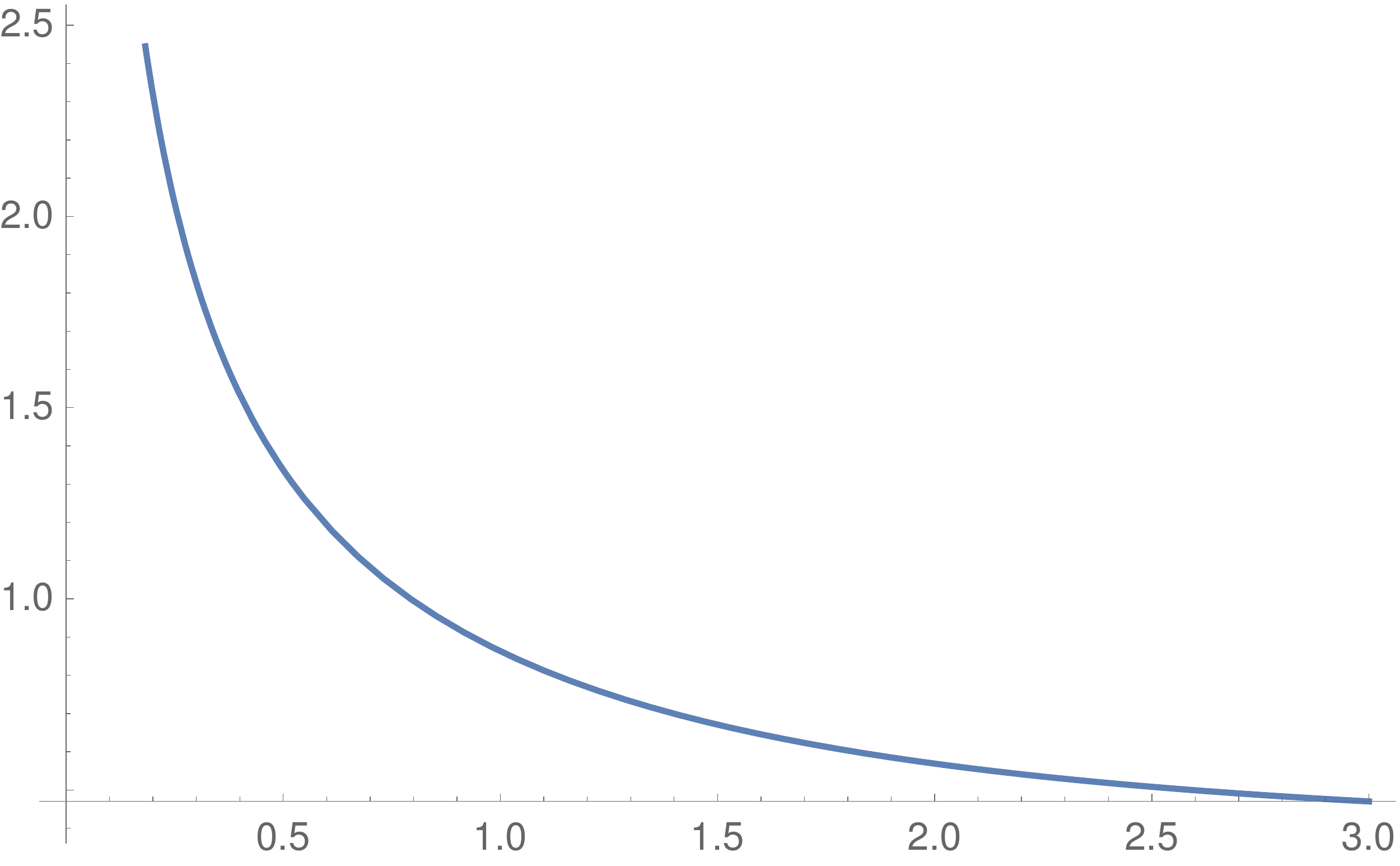}
\end{figure*}

\begin{figure*}[!htb]
\centering
\caption{The curve of acceleration. The point $ t_{c} = (3 \, ln 5 /2)  \,\sqrt{\xi}$ is the moment of passage to the positive acceleration.}
\includegraphics[width=.45\textwidth]{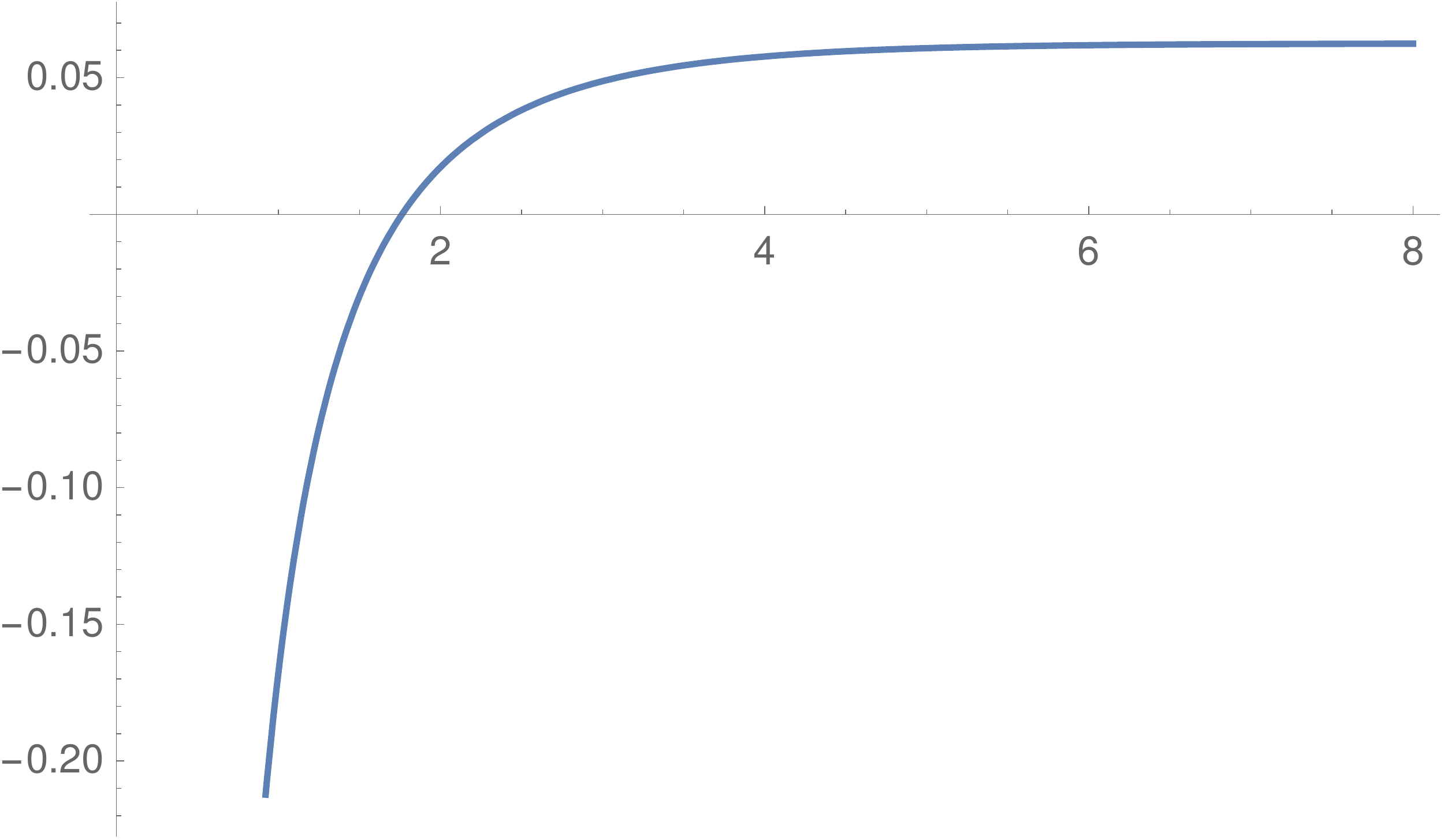}
\end{figure*}

\section{Final Comments}

In this paper we present a new way to describe the origin of the cosmological era of accelerated expansion. We show that such acceleration may be associated to a late effect of the non minimal coupling between the electromagnetic field with gravity in the realm of general relativity. A few remarks are of interest. First, we recognize that for small times, the effect of the non minimal coupling can be neglected. This means that the net effect of the electromagnetic fluid at primordial times, is to provide an expansion such that $ a(t) \approx \sqrt{t}$, that is, the scalar of curvature is approximatively null. However, for late times, when the constant $ \xi $ cannot be neglected, the value of $R$ is constant that is $ R = 1/2\xi.$ It is precisely this fact that will provide for late times the presence of an effective cosmological constant, that is hidden in the combined energy-momentum tensor $ Z_{\mu\nu}.$ Indeed in the limit for late values of time it follows that $R_{\mu\nu} - 1/2 \, R\, g_{\mu\nu} = \Lambda \, g_{\mu\nu} $ where $ \Lambda = -  1/8\xi. $ A simple inspection on the acceleration $ \ddot{a}/a$ shows that it changes sign at the point  $ t_{c} = (3 \, ln 5 \,\sqrt{\xi})/2.$ It remains the task to develop the comparison of our present model with recent observations. This is a work in progress.

\subsection*{Acknowledgements}

We would like to acknowledge the financial support from brazilian agencies Finep, Capes, Faperj and CNPq.


\begin{thebibliography}{100}


\bibitem{nb} See M Novello and S E P Bergliaffa in Bouncing Cosmologies ( Physics Reports 463 (2008) 127) for a review on non minimal coupling in cosmology and others references therein.


\bibitem{de} For a review on the subject and others references see Varun Sahni \emph{Dark Energy} in XIth Brazilian School of Cosmology and Gravitation (American Institute of Physics. Eds. M Novello and S Bergliaffa, 2005).

\bibitem{tolmanbook} R. Tolman \emph{Relativity, Thermodynamics, and Cosmology}, Oxford, Clarendon (1934).
\bibitem{anysigma} In order to obtain the expressions for $a(t)$ and $\theta$ for other values of $ \sigma$ it is enough to make the substitution $ \xi \rightarrow \xi \, (\sigma^{2} + 1).$
  \end{thebibliography}
\end{document}